\documentstyle[prl,aps,epsf]{revtex}

\begin{document}

\title{Magnetic field induced polarization effects in intrinsically granular superconductors}

\author{S. Sergeenkov$^{a,b}$}

\address{$^{a}$Centro de F\'\i sica das Interac\c c\~oes Fundamentais, Instituto Superior
T\'ecnico,\\
Av. Rovisco Pais, 1049-001 Lisboa, Portugal\\ $^{b}$Laboratory
of Theoretical Physics, Joint Institute for Nuclear Research,\\
141980 Dubna, Moscow Region, Russia}

\date{revised version; \today}
\maketitle

\begin{abstract}
Based on the previously suggested model of nanoscale dislocations
induced Josephson junctions and their arrays, we study the
magnetic field induced electric polarization effects in
intrinsically granular superconductors. In addition to a new
phenomenon of chemomagnetoelectricity, the model predicts also a
few other interesting effects, including charge analogues of
Meissner paramagnetism (at low fields) and "fishtail" anomaly (at
high fields). The conditions under which these effects can be
experimentally measured in non-stoichiometric high-$T_c$
superconductors are discussed.
\end{abstract}


\section{Introduction}

Both granular superconductors and artificially prepared arrays of
Josephson junctions (JJAs) proved useful in studying the numerous
quantum (charging) effects in these interesting systems, including
blockade of Cooper pair tunneling~\cite{1}, Bloch
oscillations~\cite{2}, propagation of quantum ballistic
vortices~\cite{3}, spin-tunneling related effects using specially
designed $SFS$-type junctions~\cite{4,5}, novel Coulomb effects in
$SINIS$-type nanoscale junctions~\cite{6}, and recently observed
geometric quantization phenomena~\cite{7} (see, e.g.,
Ref.~\cite{8} for the recent review on charge and spin effects in
mesoscopic 2D Josephson junctions).

More recently, it was realized that JJAs can be also used as
quantum channels to transfer quantum information between distant
sites~\cite{9,10,11} through the implementation of the so-called
superconducting qubits which take advantage of both charge and
phase degrees of freedom (see, e.g., Ref.~\cite{12} for a review
on quantum-state engineering with Josephson-junction devices).

At the same time, imaging of the granular structure in underdoped
$Bi_2Sr_2CaCu_2O_{8+\delta}$ crystals~\cite{13}, revealed an
apparent charge segregation of its electronic structure into
superconducting domains (of the order of a few nanometers) located
in an electronically distinct background. In particular, it was
found that at low levels of hole doping ($\delta \le 0.2$), the
holes become concentrated at certain hole-rich domains. Tunneling
between such domains leads to intrinsic granular superconductivity
(GS) in high-$T_c$ superconductors (HTS). As was shown
earlier~\cite{14}, GS based phenomena can shed some light on the
origin and evolution of the so-called paramagnetic Meissner effect
(PME) which manifests itself both in high-$T_c$ and conventional
superconductors~\cite{15,16}.

In this paper, within a previously suggested~\cite{14} model of
JJAs created by a regular 2D network of twin-boundary (TB)
dislocations with strain fields acting as an insulating barrier
between hole-rich domains in underdoped crystals, we address yet
another class of interesting phenomena which are actually dual to
the chemomagnetic effects described in Ref.~\cite{14}.
Specifically, we discuss a possible existence of a non-zero
electric polarization $P(B,\delta )$ (chemomagnetoelectic effect)
and the related change of the charge balance in intrinsically
granular non-stoichiometric material under the influence of an
applied magnetic field. In particular, we predict an anomalous
low-field magnetic behavior of the effective junction charge
$Q(B,\delta )$ and concomitant magnetocapacitance $C(B,\delta )$
in paramagnetic Meissner phase and a charge analog of
"fishtail"-like anomaly at high magnetic fields.

\section{The model}

Recall that the observed~\cite{13,17,18,19,20} in HTS single
crystals regular 2D dislocation networks of oxygen depleted
regions with the size $d_0$ of a few Burgers vectors can provide
quite a realistic possibility for existence of 2D Josephson
network within $CuO$ plane~\cite{21,22}. In this regard, it is
also important to mention the pioneer works by Khaikin and
Khlyustikov~\cite{23,24,25} on twinning-induced superconductivity
in dislocated crystals.

At the same time, in underdoped crystals there is a real
possibility to facilitate oxygen transport via the so-called
osmotic mechanism~\cite{14,19,20,26} which relates a local value
of the chemical potential $\mu ({\bf x})=\mu (0)+\nabla \mu \cdot
{\bf x}$ with a local concentration of point defects as follows
$c({\bf x})=e^{-\mu ({\bf x})/k_BT}$, and allows us to explicitly
incorporate the oxygen deficiency parameter $\delta $ into our
model by relating it to the excess oxygen concentration of
vacancies $c_v\equiv c(0)$ as follows $\delta=1-c_v$. Assuming the
following connection between the variation of mechanical and
chemical properties of planar defects, namely $\mu ({\bf
x})=K\Omega _0\epsilon ({\bf x})$ (where $\epsilon ({\bf x})=
\epsilon _0e^{-{\mid{{\bf x}}\mid}/d_0}$ is screened strain field
produced by tetragonal regions in $d$-wave orthorhombic $YBCO$
crystal, $\Omega _0$ is an effective atomic volume of the vacancy,
and $K$ is the bulk elastic modulus), we can study the properties
of TB induced JJs under intrinsic chemical pressure $\nabla \mu$
(created by the variation of the oxygen doping parameter $\delta
$). More specifically, a single $SIS$ type junction (comprising a
Josephson network) is formed around TB due to a local depression
of the superconducting order parameter $\Delta ({\bf x})\propto
\epsilon({\bf x})$ over distance $d_0$ producing thus a weak link
with Josephson coupling $J(\delta )=\epsilon({\bf
x})J_0=J_0(\delta )e^{-{\mid{{\bf x}}\mid}/d_0}$ where $J_0(\delta
)=\epsilon _0J_0=(\mu _v/K\Omega _0 )J_0$ (here $J_0\propto \Delta
_0/R_n$ with $R_n$ being a resistance of the junction). Notice
that, in accordance with the observations, for stoichiometric
situation (when $\delta \simeq 0$), the Josephson coupling
$J(\delta ) \simeq 0$ and the system loses its explicitly granular
signature.

To describe the influence of chemomagnetic effects on charge
balance of an intrinsically granular superconductor, we employ a
model of 2D overdamped Josephson junction array which is based on
the well known Hamiltonian
\begin{equation}
{\cal H}=\sum_{ij}^NJ_{ij}(1-\cos \phi_{ij})+\sum_{ij}^N
\frac{q_iq_j}{2C_{ij}}
\end{equation}
and introduces a short-range (nearest-neighbor) interaction between
$N$ junctions (which are formed around oxygen-rich superconducting
areas with phases $\phi _i$), arranged in a two-dimensional (2D)
lattice with coordinates ${\bf x_i}=(x_i,y_i)$. The areas are
separated by oxygen-poor insulating boundaries (created by TB strain
fields $\epsilon({\bf x}_{ij})$) producing a short-range Josephson
coupling $J_{ij}=J_0(\delta )e^{-{\mid{{\bf x}_{ij}}\mid}/d}$. Thus,
typically for granular superconductors, the Josephson energy of the
array varies exponentially with the distance ${\bf x}_{ij}={\bf
x}_{i}-{\bf x}_{j}$ between neighboring junctions (with $d$ being an
average junction size). As usual, the second term in the rhs of Eq.(1)
accounts for Coulomb effects where $q_i =-2en_i$ is the junction charge
with $n_i$ being the pair number operator. Naturally, the same strain
fields $\epsilon({\bf x}_{ij})$ will be responsible for dielectric properties
of oxygen-depleted regions as well via the $\delta $-dependent capacitance tensor
$C_{ij}(\delta )=C[\epsilon({\bf x}_{ij})]$.

If, in addition to the chemical pressure $\nabla \mu ({\bf
x})=K\Omega _0\nabla \epsilon ({\bf x})$, the network of
superconducting grains is under the influence of an applied
frustrating magnetic field ${\bf B}$, the total phase difference
through the contact reads
\begin{equation}
\phi _{ij}=\phi ^0_{ij}+\frac{\pi w}{\Phi _0} ({\bf x}_{ij}\wedge
{\bf n}_{ij})\cdot {\bf B}+\frac{\nabla \mu \cdot {\bf
x}_{ij}t}{\hbar},
\end{equation}
where $\phi ^0_{ij}$ is the initial phase difference (see below),
${\bf n}_{ij}={\bf X}_{ij}/{\mid{{\bf X}_{ij}}\mid}$ with $ {\bf
X}_{ij}=({\bf x}_{i}+{\bf x}_{j})/2$, and $w=2\lambda _L(T)+l$
with $\lambda _L$ being the London penetration depth of
superconducting area and $l$ an insulator thickness (which, within
the discussed here scenario, is simply equal to the TB
thickness~\cite{26}).

As usual, to safely neglect the influence of the self-field
effects in a real material, the corresponding Josephson
penetration length $\lambda _J=\sqrt{\Phi _0/2\pi \mu _0j_c w}$
must be larger than the junction size $d$. Here $j_c$ is the
critical current density of superconducting (hole-rich) area. As
we shall see below, this condition is rather well satisfied for
HTS single crystals.

\section{Chemomagnetoelectricity}

In what follows, we shall be interested in the behavior of
magnetic field induced electric polarization
(chemomagnetoelectricity) in chemically induced GS described by a
$2D$ JJA. Recall that a conventional (zero-field) pair
polarization operator within the model under discussion
reads~\cite{27,28}
\begin{equation}
{\bf p}=\sum_{i=1}^Nq_i {\bf x}_{i}
\end{equation}
In view of Eqs.(1)-(3), and taking into account a usual
"phase-number" commutation relation, $[\phi _i,n_j]=i\delta
_{ij}$, it can be shown that the evolution of the pair polarization operator is
determined via the equation of motion
\begin{equation}
\frac{d{\bf p}}{dt}=\frac{1}{i\hbar}\left[ {\bf p},{\cal H}\right
] =\frac{2e}{\hbar }\sum_{ij}^NJ_{ij}\sin \phi _{ij}(t){\bf x}_{ij}
\end{equation}
Resolving the above equation, we arrive at the following net value
of the magnetic-field induced longitudinal (along $x$-axis)
electric polarization ${P}(\delta ,{\bf B}) \equiv <{p}_x(t)>$
and the corresponding effective junction charge
\begin{equation}
{Q}(\delta ,{\bf B})=\frac{2eJ_0} {\hbar \tau d} \int\limits_{0}^
{\tau }dt \int \limits_{0}^{t}dt'\int \frac{d^2x}{S}
\sin \phi ({\bf x}, t')xe^{-{\mid{{\bf x}}\mid}/d},
\end{equation}
where $S=2\pi d^2$ is properly defined normalization area, $\tau$
is a characteristic time (see Discussion), and we made a usual
substitution $\frac{1}{N}\sum_{ij}A_{ij}(t) \to \frac{1}{S}\int
d^2x A({\bf x},t)$ valid in the long-wavelength
approximation~\cite{28}.

To capture the very essence of the superconducting analog of the
chemomagnetoelectric effect, in what follows we assume for
simplicity that a {\it stoichiometric sample} (with $\delta \simeq
0$) does not possess any spontaneous polarization at zero magnetic
field, that is $P(0,0)=0$. According to Eq.(5), this condition
implies $\phi _{ij}^0=2\pi m$ for the initial phase difference
with $m=0,\pm 1, \pm 2,..$.

Taking the applied magnetic field along the $c$-axis (and normal to
the $CuO$ plane), that is ${\bf B}=(0,0,B)$, we obtain finally
\begin{equation}
Q(\delta ,B)=Q_0(\delta ) \frac{2{\tilde b}+b(1-{\tilde
b}^2)}{(1+b^2)(1+{\tilde b}^2)^2}
\end{equation}
for the magnetic field behavior of the effective junction charge in chemically
induced granular superconductors.

Here $Q_0(\delta )=e\tau J_0(\delta )/\hbar$ with $J_0(\delta )$
defined earlier, $b=B/B_0$, ${\tilde b}=b-b_{\mu }$, and $b_{\mu
}=B_{\mu }/B_0\simeq (k_BT\tau /\hbar )\delta $ where $B_{\mu
}(\delta )=(\mu _v\tau /\hbar )B_0$ is the chemically-induced
contribution (which disappears in optimally doped systems with
$\delta \simeq 0$), and $B_0=\Phi _0/wd$ is a characteristic
Josephson field.

Fig.1 shows changes of the initial (stoichiometric) effective
junction charge $Q$ (solid line) with oxygen deficiency $\delta$.
Notice a sign change of $Q$ (dotted and dashed lines) driven by
non-zero values of $\delta$ at low magnetic fields (a charge
analog of chemically induced PME). According to Eq.(6), the
effective charge changes its sign as soon as the chemomagnetic
contribution $B_{\mu }(\delta )$ exceeds an applied magnetic field
$B$ (see Discussion).

At the same time, Fig.2 presents a true {\it chemoelectric} effect
with concentration (deficiency) induced effective junction charge
$Q(\delta ,0)$ in zero magnetic field. Notice that $Q(\delta ,0)$
exhibits a maximum around $\delta _c\simeq 0.2$ (in agreement with
the classical percolative behavior observed in non-stoichiometric
$YBa_2Cu_3O_{7-\delta }$ samples~\cite{17}).

It is of interest also to consider the magnetic field behavior of
the concomitant effective flux capacitance $C\equiv \tau dQ(\delta ,B)/d\Phi $
which in view of Eq.(6) reads
\begin{equation}
C(\delta ,B)=C_0(\delta )\frac{1-3b{\tilde b}-3{\tilde
b}^2+b{\tilde b}^3}{(1+b^2)(1+{\tilde b}^2)^3},
\end{equation}
where $\Phi =SB$, and $C_0(\delta )=\tau Q_0(\delta )/\Phi _0$.

Fig.3 depicts the behavior of the effective flux capacitance
$C(\delta ,B)$ in applied magnetic field for different values of
oxygen deficiency parameter: $\delta \simeq 0$ (solid line),
$\delta =0.1$ (dashed line), and $\delta=0.2$ (dotted line).
Notice a decrease of magnetocapacitance amplitude and its peak
shifting with increase of $\delta$ and sign change at low magnetic
fields which is another manifestation of the charge analog of
chemically induced PME (Cf. Fig.1).

\section{Charge analog of "fishtail" anomaly}

So far, we neglected a possible field dependence of the chemical
potential $\mu _v$ of oxygen vacancies. Recall, however, that in
high enough applied magnetic fields $B$, the field-induced change
of the chemical potential $\Delta \mu _v(B)\equiv \mu _v(B)-\mu
_v(0)$ becomes tangible and should be taken into
account~\cite{14,29,30}. As a result, we end up with a
superconducting analog of the so-called {\it magnetoconcentration}
effect~\cite{14} with field induced creation of oxygen vacancies
$c_v(B)=c_v(0)\exp(-\Delta \mu _v(B)/k_BT)$ which in turn brings
about a "fishtail"-like behavior of the high-field
chemomagnetization (see Ref.~\cite{14} for more details).

Fig.4 shows the field behavior of the effective junction charge in
the presence of the above-mentioned magnetoconcentration effect.
As it is clearly seen, $Q(\delta (B),B)$ exhibits a
"fishtail"-like anomaly typical for previously discussed~\cite{14}
chemomagnetization in underdoped crystals with intragrain
granularity (for symmetry and better visual effect we also plotted
$-Q(\delta (B),B)$ in the same figure). This more complex
structure of the effective charge appears when the applied
magnetic field $B$ matches an intrinsic chemomagnetic field
$B_{\mu}(\delta (B))$ (which now also depends on $B$ via the
magnetoconcentration effect). Notice that a "fishtail" structure
of $Q(\delta (B),B)$ manifests itself even at zero values of
field-free deficiency parameter $\delta (0)$ (solid line in Fig.4)
thus confirming a field-induced nature of intrinsic
granularity~\cite{13,17,18,19,20}. Likewise, Fig.5 depicts the
evolution of the effective flux capacitance $C(\delta (B),B)$ in
applied magnetic field $B/B_0$ in the presence of
magnetoconcentration effect (Cf. Fig.3).

\section{Discussion}

Thus, the present model predicts appearance of two interrelated
phenomena (dual to the previously discussed behavior of
chemomagnetizm~\cite{14}), namely a charge analog of Meissner
paramagnetism at low fields and a charge analog of "fishtail"
anomaly at high fields. To see whether these effects can be
actually observed in a real material, let us estimate an order of
magnitude of the main model parameters.

Using typical~\cite{17,19} for HTS single crystals values of
$\lambda _L(0) \simeq 150nm$, $d \simeq 10nm$, and $j_c \simeq
10^{10}A/m^2$, we arrive at the following estimates of the
characteristic $B_0 \simeq 0.5T$  and chemomagnetic $B_{\mu
}(\delta ) \simeq 0.5B_0$ fields, respectively. So, the predicted
charge analog of PME should be observable for applied magnetic
fields $B < 0.25T$. Notice that, for the above set of parameters,
the Josephson length is of the order of $\lambda _J \simeq 1\mu
m$, which means that the assumed in this paper small-junction
approximation is valid and the "self-field" effects can be safely
neglected.

Furthermore, the characteristic frequencies $\omega \simeq \tau
^{-1}$ needed to probe the suggested here effects are related to
the processes governed by tunneling relaxation times $\tau \simeq
\hbar /J_0(\delta )$. Since for oxygen deficiency parameter
$\delta =0.1$ the chemically-induced zero-temperature Josephson
energy in non-stoichiometric $YBCO$ single crystals is of the
order of $J_0(\delta ) \simeq k_B T_C \delta \simeq 1meV$, we
arrive at the required frequencies of $\omega \simeq 10^{13}Hz$
and at the following estimates of the effective junction charge
$Q_0 \simeq e=1.6\times 10^{-19}C$ and flux capacitance $C_0
\simeq 10^{-18}F$. Notice that the above estimates fall into the
range of parameters used in typical experiments for studying the
single-electron tunneling effects both in JJs and
JJAs~\cite{1,2,12,31} suggesting thus quite an optimistic
possibility to observe the above-predicted field induced effects
experimentally in non-stoichiometric superconductors with
pronounced networks of planar defects or in artificially prepared
JJAs. (It is worth mentioning that a somewhat similar behavior of
the magnetic field induced charge and related flux capacitance has
been observed in 2D electron systems~\cite{32}.)

And finally, it can be easily verified that, in view of
Eqs.(1)-(5), the field-induced Coulomb energy of the
oxygen-depleted region within our model is given by
\begin{equation}
E_C(\delta ,B)\equiv \left < \sum_{ij}^N \frac{q_iq_j}{2C_{ij}}
\right >=\frac{Q^2(\delta ,B)}{2C(\delta ,B)}
\end{equation}
with $Q(\delta ,B)$ and $C(\delta ,B)$ defined by Eqs. (6) and
(7), respectively.

A thorough analysis of the above expression reveals that in the
PME state (when $B\ll B_{\mu }$) the chemically-induced granular
superconductor is always in the so-called Coulomb blockade regime
(with $E_C>J_0$), while in the ''fishtail'' state (for $B\ge
B_{\mu }$) the energy balance tips in favor of tunneling (with
$E_C<J_0$). In particular, $E_C(\delta ,B=0.1B_{\mu
})=\frac{\pi}{2}J_0(\delta )$ and $E_C(\delta ,B=B_{\mu
})=\frac{\pi}{8}J_0(\delta )$. It would be also interesting to
check this phenomenon of field-induced weakening of the Coulomb
blockade experimentally.

\section{Conclusion}

In conclusion, within a realistic model of 2D Josephson junction
arrays created by 2D network of twin boundary dislocations (with
strain fields acting as an insulating barrier between hole-rich
domains in underdoped crystals), a few novel electric polarization
related effects expected to occur in intrinsically granular
material under applied magnetic fields were predicted, including a
phenomenon of chemomagnetoelectricity, an anomalous low-field
magnetic behavior of the effective junction charge (and flux
capacitance) in paramagnetic Meissner phase and a charge analog of
"fishtail"-like anomaly at high magnetic fields  as well as
field-dependent weakening of the chemically-induced Coulomb
blockade. The experimental conditions needed to observe the
predicted here effects in non-stoichiometric high-$T_c$
superconductors were discussed.

\acknowledgments

 This work was done during my stay at the Center
for Physics of Fundamental Interactions (Instituto Superior
T\'ecnico, Lisbon) and was partially funded by the FCT. I thank
Pedro Sacramento and Vitor Vieira for hospitality and interesting
discussions on the subject. I am also indebted to the Referee for
drawing my attention to the pioneer works by Khaikin and
Khlyustikov on twinning-induced superconductivity in dislocated
crystals.

\begin{figure}[tbh]
\epsfxsize=12.5cm \centerline{\hspace{0mm} \epsffile{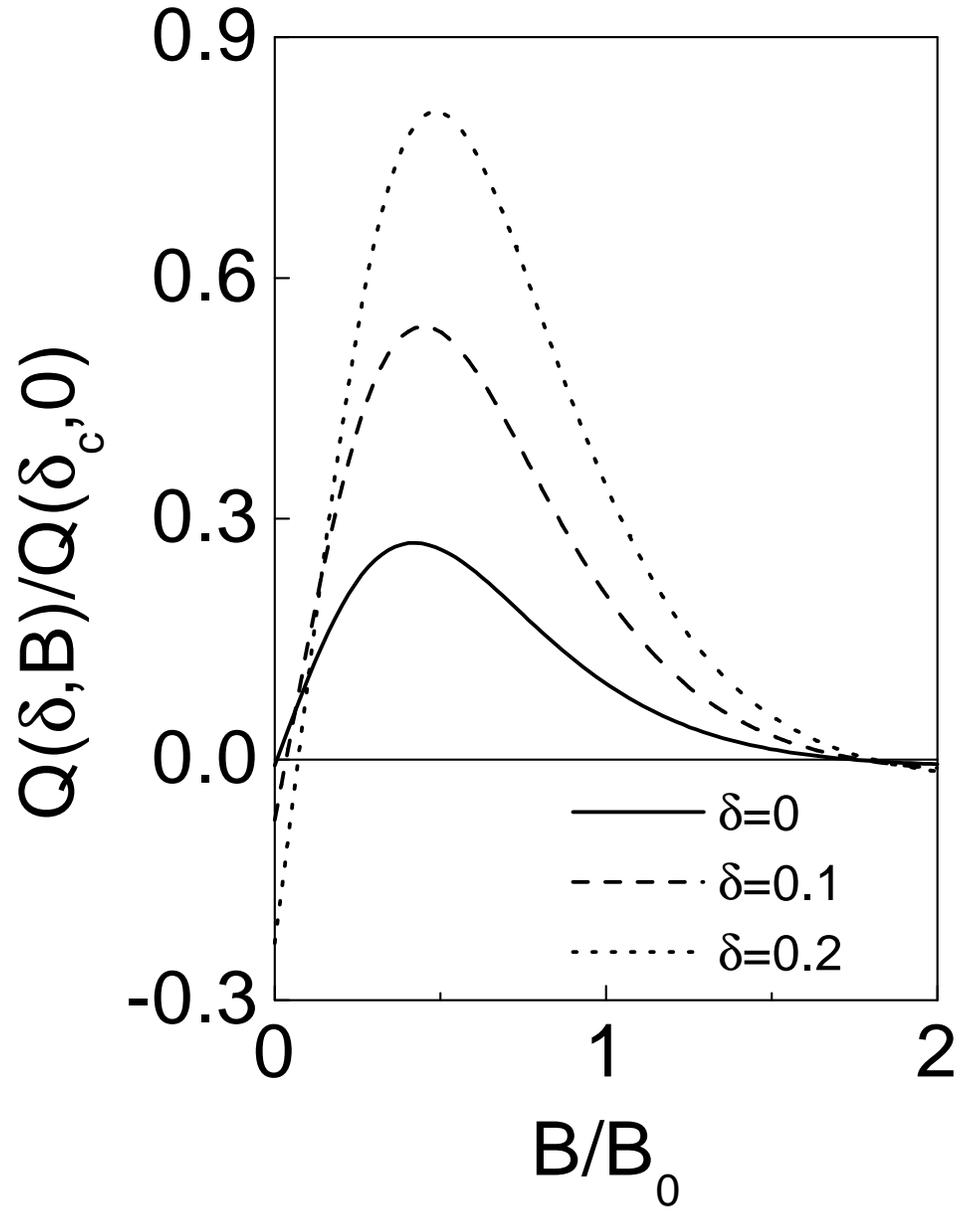}}
\vspace{2cm} \caption{The effective junction charge $Q(\delta
,B)/Q(\delta _c,0)$ (chemomagnetoelectric effect) as a function of
applied magnetic field $B/B_0$, according to
 Eq.(6), for different values of oxygen deficiency parameter: $\delta \simeq 0$ (solid line),
 $\delta =0.1$ (dashed line), and $\delta=0.2$ (dotted line). }
\end{figure}
\newpage
\begin{figure}[tbh]
\epsfxsize=12.5cm \centerline{\hspace{0mm} \epsffile{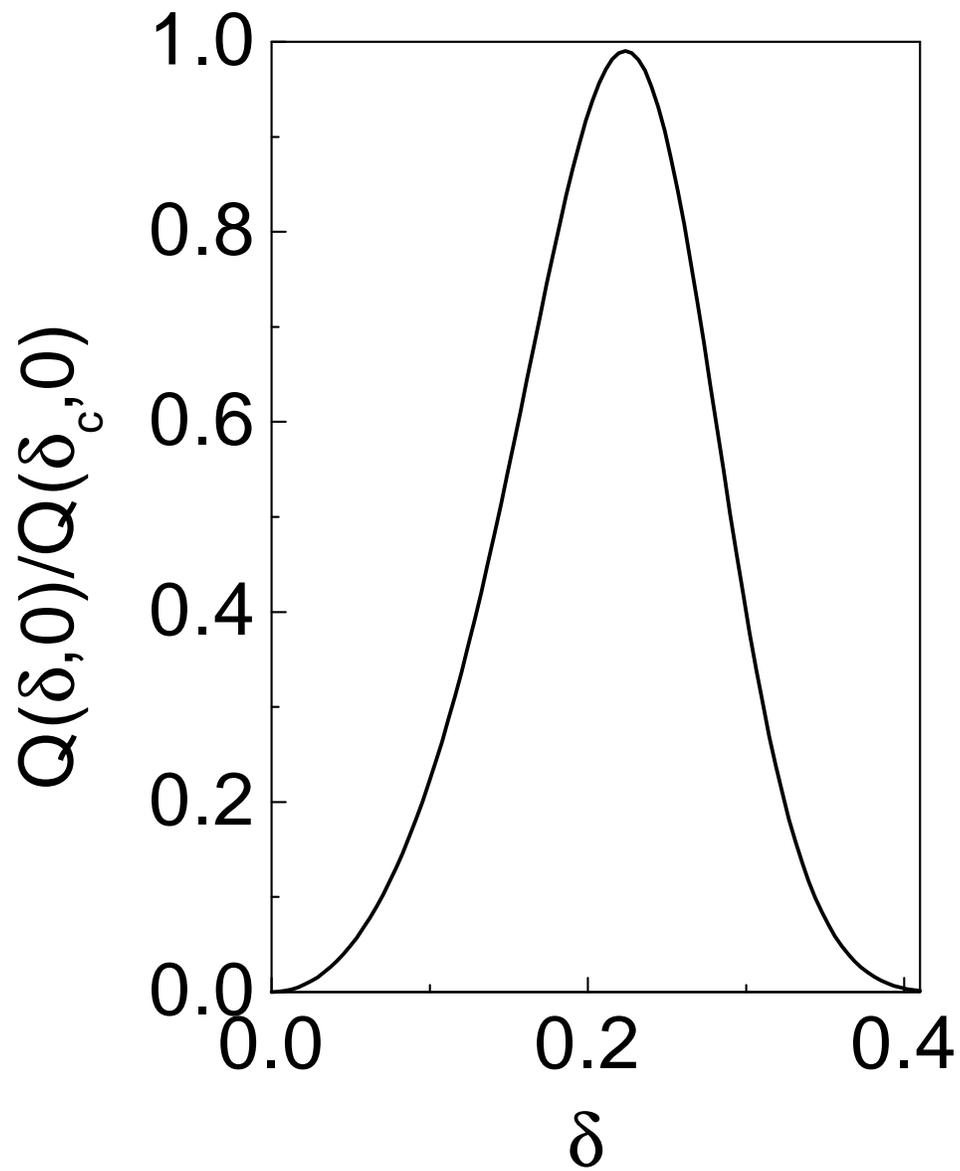}}
\vspace{2cm} \caption{Chemically induced effective junction charge
$Q(\delta ,0)/Q(\delta _c,0)$ in a zero applied magnetic field
(true chemoelectric effect). }
\end{figure}
\newpage
\begin{figure}[tbh]
\epsfxsize=12.5cm \centerline{\hspace{0mm} \epsffile{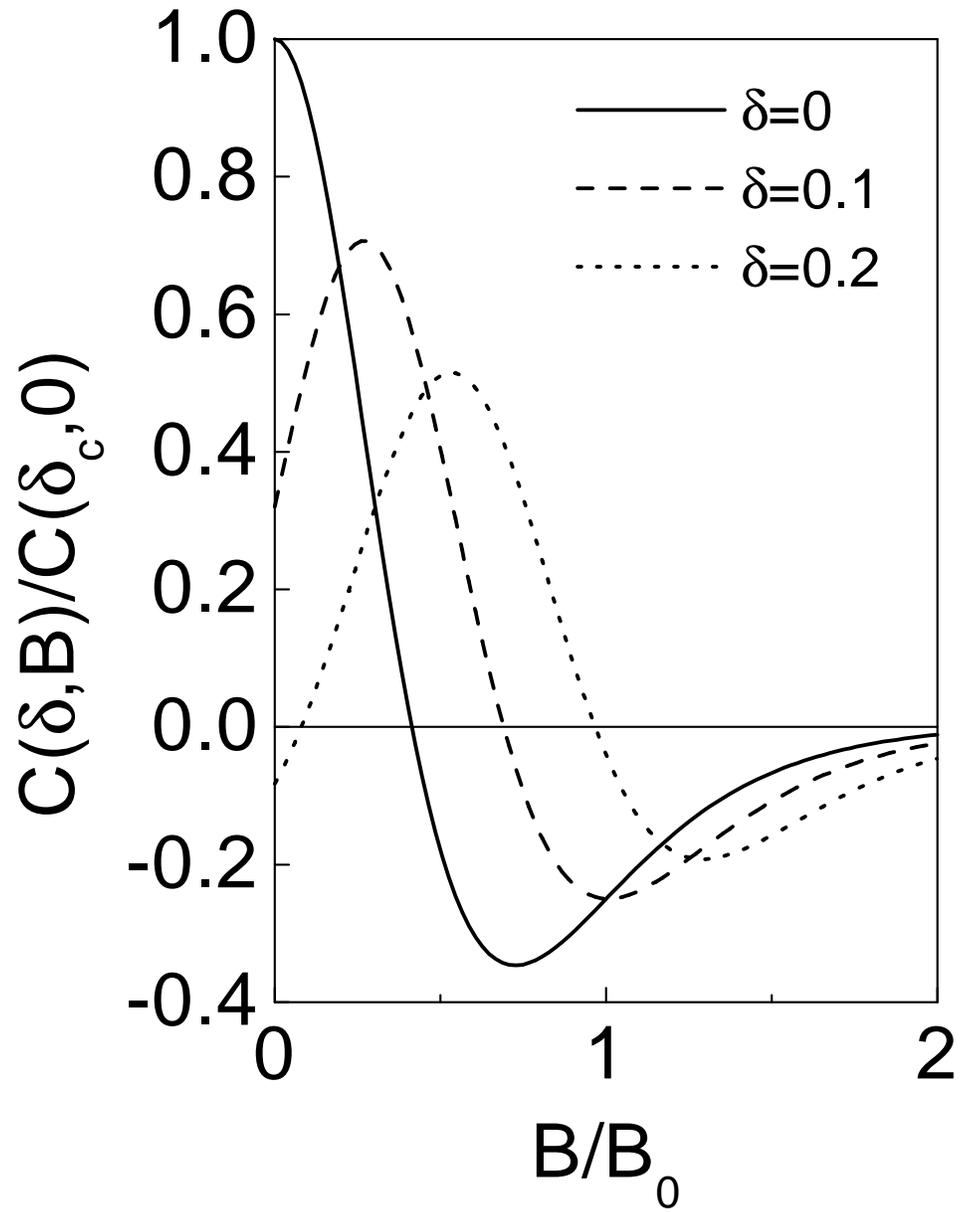}}
\vspace{2cm} \caption{The effective flux capacitance $C(\delta
,B)/C(\delta _c, 0)$
 as a function of applied magnetic field $B/B_0$, according to Eq.(7),
 for different values of oxygen deficiency parameter: $\delta \simeq 0$ (solid line),
 $\delta =0.1$ (dashed line), and $\delta=0.2$ (dotted line).}
\end{figure}
\newpage
\begin{figure}[tbh]
\epsfxsize=12.5cm \centerline{\hspace{0mm} \epsffile{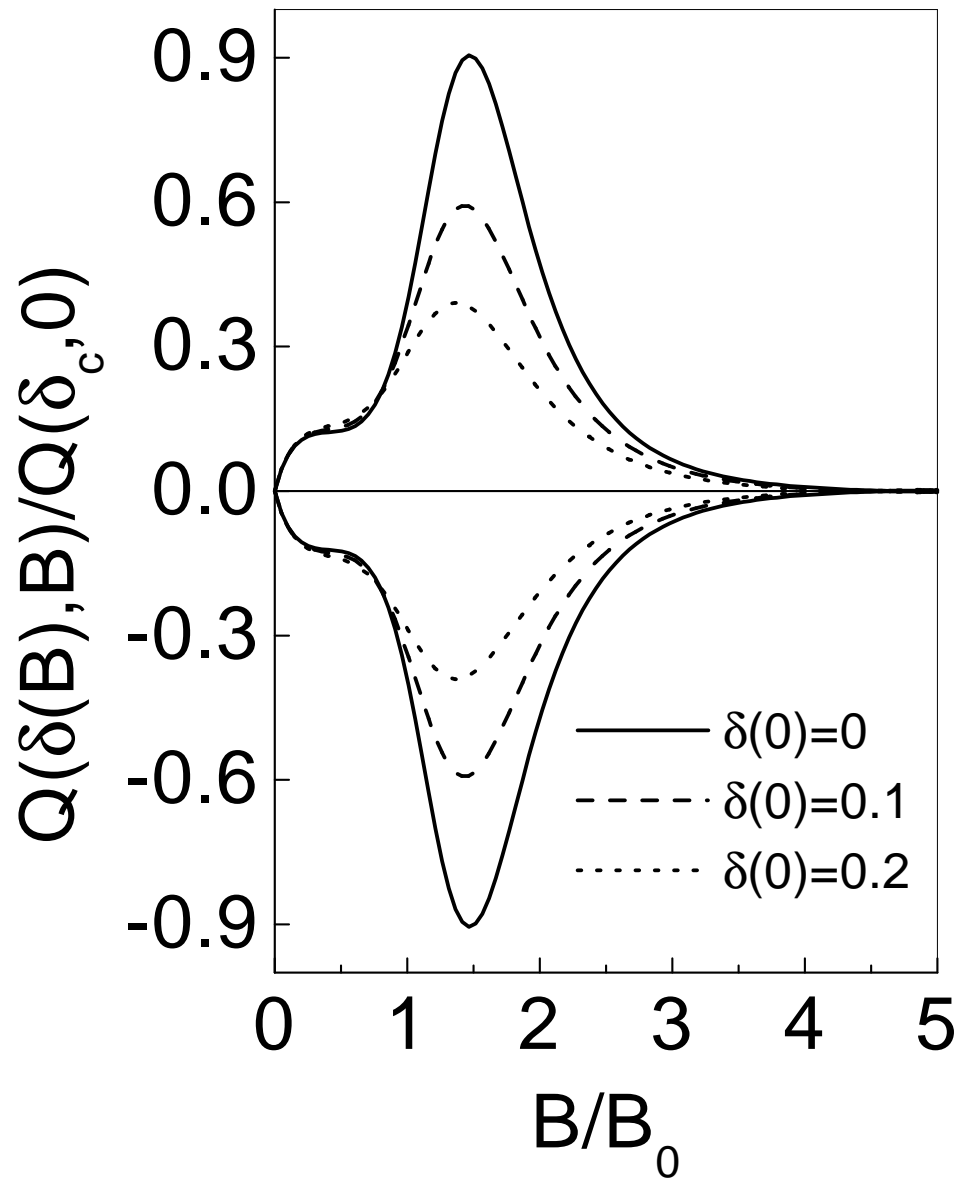}}
\vspace{2cm} \caption{A "fishtail"-like behavior of effective
charge $Q(\delta (B),B)/Q(\delta _c,0)$ in applied magnetic field
$B/B_0$ in the presence
 of magnetoconcentration effect (with field-induced oxygen vacancies
 $\delta (B)$) for three values of field-free deficiency parameter:
 $\delta (0)\simeq 0$ (solid line), $\delta (0)=0.1$ (dashed line), and $\delta (0)=0.2$ (dotted line).}
\end{figure}
\newpage
\begin{figure}[tbh]
\epsfxsize=12.5cm \centerline{\hspace{0mm} \epsffile{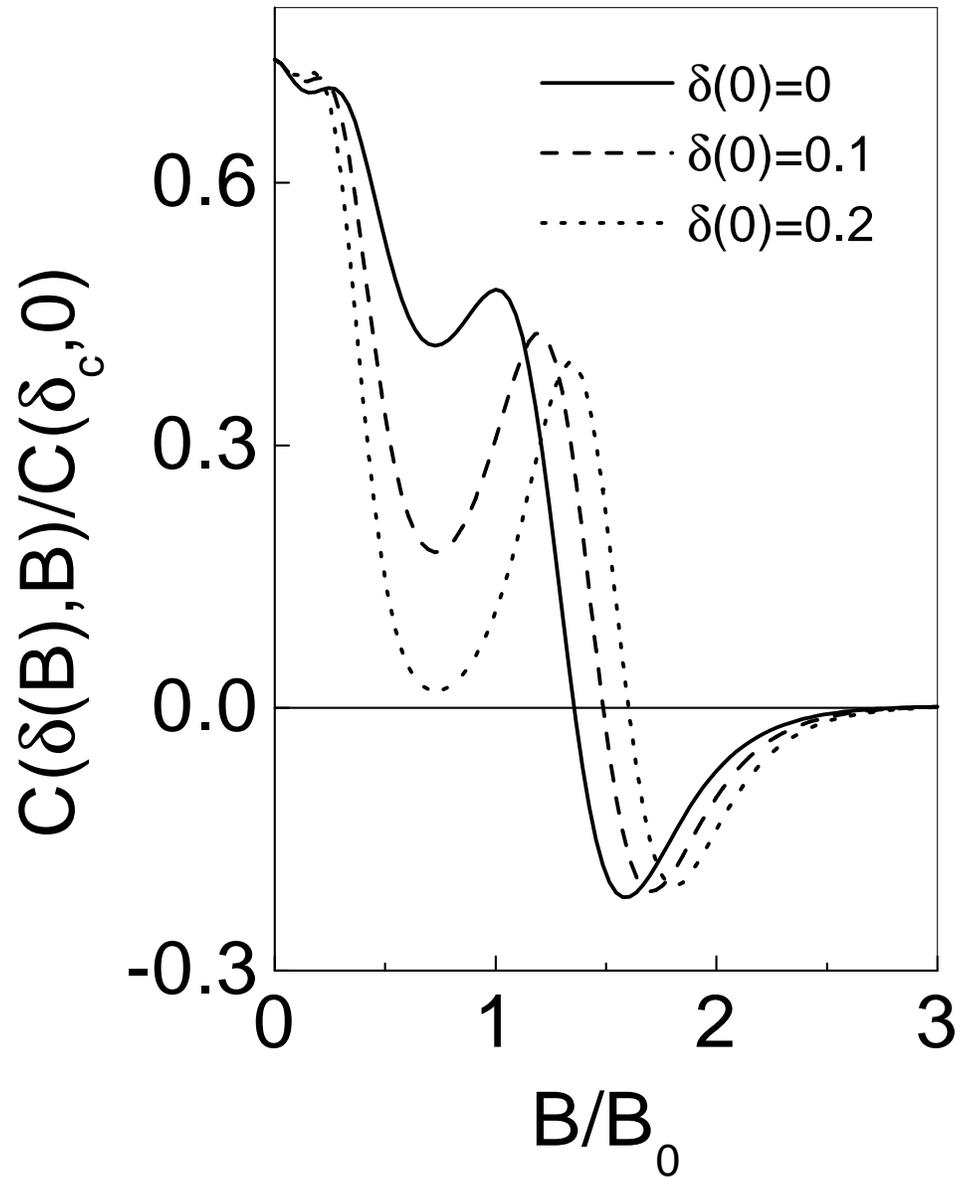}}
\vspace{2cm} \caption{The behavior of the effective flux
capacitance
 $C(\delta (B),B)/C(\delta _c, 0)$ in applied magnetic field $B/B_0$ in the presence
 of magnetoconcentration effect for three values of field-free deficiency parameter:
 $\delta (0)\simeq 0$ (solid line), $\delta (0)=0.1$ (dashed line), and $\delta (0)=0.2$ (dotted line).}
\end{figure}

\end{document}